\documentclass[preprint,aip,jcp,groupaddress,floatfix]{revtex4-1}

\usepackage{graphicx}
\usepackage{bm}
\usepackage{color}
\usepackage{amsmath}
\usepackage{amssymb}
\usepackage{dsfont}
\usepackage{color}
\usepackage{calc}
\usepackage{hyperref}
\usepackage{multirow}
\usepackage{natmove}
\usepackage{wasysym}

\graphicspath{{./figures/}}

\newcommand{\lla}{\left\langle}
\newcommand{\rra}{\right\rangle}
\newcommand{\psum}{\sideset{}{'}\sum}
\newcommand{\brn}{{\bm R}_{\bm n}}

\begin{document}
\title{Virial pressure in systems of active Brownian particles}

\author{Roland G. Winkler}
\author{Adam Wysocki}
\author{Gerhard Gompper}
\affiliation{Theoretical Soft Matter and Biophysics, Institute for
Advanced Simulation and Institute of Complex Systems,
Forschungszentrum J\"ulich, D-52425 J\"ulich, Germany}

\date{\today}

\begin{abstract}
The pressure of suspensions of self-propelled objects is studied theoretically and by simulation of spherical active Brownian particles (ABP). We show that for certain geometries, the mechanical pressure as force/area of a confined systems can equally be expressed by bulk properties, which implies the existence of an nonequilibrium equation of state. Exploiting the virial theorem, we derive expressions for the pressure of ABPs confined by solid walls or exposed to periodic boundary conditions. In both cases, the
pressure comprises three contributions: the ideal-gas pressure due to white-noise random forces, an activity-induce pressure (''swim pressure''), which can be expressed in terms of a product of the bare and a mean effective propulsion velocity,  and the contribution by interparticle forces. We find that the pressure of spherical ABPs in confined systems explicitly depends on the presence of the confining walls and the particle-wall interactions, which has no correspondence in systems with periodic boundary conditions. Our simulations of three-dimensional APBs in systems with periodic boundary conditions reveal a pressure-concentration dependence that becomes increasingly nonmonotonic with increasing activity. Above a critical activity and ABP concentration, a phase transition occurs, which is reflected in a rapid and steep change of the pressure. We present and discuss the pressure for various activities and analyse the contributions of the individual pressure components.
\end{abstract}
\maketitle

\section{Introduction}

Living matter composed of active particles, which convert internal energy into systematic translational motion, is a particular class of materials typically far from equilibrium. Examples range from the microscopic scale of bacterial suspensions\cite{elge:15} to the macroscopic scale of flocks of birds and mammalian herds.\cite{vics:12} Such active systems exhibit remarkable nonequilibrium phenomena and emergent behavior like swarming, \cite{cope:09,darn:10,kear:10,dres:11,part:13} turbulence, \cite{dres:11} activity-induced clustering and phase transitions, \cite{bial:12,butt:13,mogn:13,theu:12,fily:14.1,yang:14.2,sten:14,fily:14,redn:13,fily:12,zoet:14,wyso:14} and a shift of the glass-transition temperature. \cite{ni:13,wyso:14} The nonequilibrium character of active matter poses particular challenges for a theoretical description. In particular, a thermodynamic description is missing, but would be desirable in order to be able to define elementary thermodynamic variables such as temperature or chemical potential. \cite{solo:15} Considerable progress in the description of activity-induced phase transitions has been achieved recently. \cite{bial:13,sten:13,witt:14,sten:14,solo:15,solo:15.1}

Pressure in active fluids attracted considerable attention lately, because a mechanical pressure as force per area can be defined far from equilibrium and might be useful in the strive to derive a nonequilibrium equation of state.\cite{taka:14,yang:14.2,gino:15,solo:15,bert:15}  Interestingly, simulation studies reveal such an equation of state, with a van der Waals-type pressure-volume phase diagram typical for gas-liquid coexistence. \cite{taka:14,yang:14.2,gino:15,solo:15}

In thermal equilibrium, pressure can equivalently be defined in various ways. In addition to the mechanical definition, pressure $p$ follows as derivative with respect to the volume $V$ of the Helmholtz free energy $F$, i.e., $p = - (\partial F/\partial V)_{T,N}$, at constant temperature $T$ and particle number $N$. \cite{beck:67} Moreover, the virial theorem can be exploited. \cite{beck:67,wink:92.1,wink:93,wink:09} These two identical formulations relate the surface pressure to bulk properties. However, for nonequilibrium systems, it is {\em a prior} not evident how to relate the surface pressure to bulk properties, as it has to be kept in mind that there is usually no free-energy functional. By now, pressure in active fluids has been calculated using various virial-type expressions, \cite{taka:14,yang:14.2,solo:15} typically without a fundamental derivation. Basic relations are provided for two-dimensional fluids in Ref.~\onlinecite{solo:15}, starting from  the mechanical definition of pressure and by exploiting the Fokker-Planck equation for the phase-space probability distribution.

Here, we apply the virial theorem to derive expressions for the pressure in active fluids. Both, fluids confined between solid walls and exposed to periodic boundary conditions are considered. As an important first step, we demonstrate that the mechanical pressure of a confined fluid can equivalently be represented by the virial of the surface forces for particular geometries such as a cuboid and a sphere. This has important implications, since it provides a relation of the mechanical pressure with bulk properties of the fluid. \cite{solo:15} Based on this expression, we derive an internal expression for the pressure, which comprises the ideal gas pressure, an swim pressure due to particle propulsion, and a contribution by the interparticle interactions.  As a result, we find that the swim pressure of the spherical ABPs explicitly depends on the presence of confining walls and the particle-wall interaction, in contrast to results presented in Ref.~\onlinecite{solo:15}, but in agreement with more general considerations of anisotropic bodies. \cite{solo:15.1,wyso:15} Systems with periodic boundary conditions naturally lack such a term. Hence, walls can induce features in confined active systems, which will not appear in periodic systems.

By Brownian dynamics simulations, we determine pressure-concentration relations for various P\'eclet numbers of three-dimensional periodic systems of active Brownian particles. As already previously observed, \cite{taka:14,yang:14.2,mall:14,solo:15} we find an increasingly nonmonotonic behavior of the pressure with increasing propulsion velocity. Above a certain velocity, we also find a van der Waals-loop-type relation. Most importantly, however, we find an abrupt change of the pressure for phase-separating fluids. Thereby, the magnitude of the pressure jump increases with increasing propulsion velocity. Hence, we argue that the nonmonotonic behavior indicates increased density fluctuations of the active fluid, but a phase transition is reflected in a more pronounced abrupt change of the pressure. A detailed analysis of the individual contributions to the pressure shows that the main effect is due to the activity of the colloidal particles, which gives rise to a so-called ''swim pressure'' as introduced in Ref.~\onlinecite{taka:14}. However, also the interparticle-force contribution exhibits an abrupt increase above the phase-transition concentrations.

The paper is organized as follow. In Sec.~\ref{sec:model}, the model of the active fluid is presented. The dynamics of ABPs in a dilute system is analysed in Sec.~\ref{sec:dynamics} for confined and unconfined particles.  Expressions for the virial pressure are derived in Sec.~\ref{sec:virial} for confined and periodic systems. Simulations results for periodic systems are presented in Sec.~\ref{sec:simulation}, and our findings are summarized in Sec.~\ref{sec:conclusions}. Details of the stochastic dynamics of the ABPs are discussed in App. \ref{app:rotation} and \ref{app:correlation}. A derivation of the equilibrium pressure via the virial theorem is presented in App.~\ref{app:equi_pressure}. Appendix~\ref{app:surf_pressure} establishes the relation between the mechanical pressure and the surface virial for active systems.

\section{Model} \label{sec:model}

We consider a system of $N$ active objects,  which are represented by hard-sphere-like particles of diameter $\sigma$.  A particle $i$ is propelled  with constant velocity $v_{0}$ along its orientation vector ${\bm e}_i$ in three dimensions (3D). \cite{bial:12,fily:12,redn:13,wyso:14} Its translational motion is described by the Langevin eqution
\begin{equation} \label{eq:BDtrans}
\dot {\bm r}_i (t)=v_{0}{\bm e}_i (t)+\frac{1}{\gamma} \left( {\bm F}_i(t)+ {\bm \Gamma}_i(t) \right),
\end{equation}
where ${\bm r}_i$ is the particle position, $\dot {\bm r}_i$ the velocity, ${\bm F}_i$ the  total force on the particle, $\gamma$ the translational friction coefficient, and ${\bm \Gamma}_i$ a Gaussian white-noise random force, with the moments
\begin{align}
\lla {\bm \Gamma}_i (t)\rra & = 0,  \\
\lla {\Gamma}_{\alpha, i}(t)  {\Gamma}_{\beta, j} (t') \rra & = 2 \gamma k_B T \delta_{\alpha \beta} \delta_{ij} \delta(t-t') .
\end{align}
Here, $T$ denotes the temperature in equilibrium, $k_B$ the Boltzmann constant, and $\alpha,  \beta \in \{x,y,x\}$. The the frication coefficient $\gamma$ is related to translational diffusion coefficient $D_t^0$ via $D_t^0 =k_BT/\gamma$. The interparticle  force is described by the repulsive (shifted) Yukawa-like potential
\begin{align} \label{eq:pot_energy}
U(r) = k_B T \left\{ \begin{array}{cc}
\displaystyle  \infty , & r \le \sigma \\[3pt] \displaystyle
u(r) - u(r_c) , & r>\sigma  \\ \displaystyle
0  , & r > r_c
\end{array}
\right. ,
\end{align}
where $u(r) = 5l e^{-(r-\sigma)/l}/(r-\sigma)$. Here,  $l$ is the interaction range and $r_c$ is the upper cut-off radius.
The orientation ${\bm e}_i$ performs a random walk according to
\begin{align} \label{eq:orient}
\dot {\bm e}_i(t)= {\bm e}_i(t) \times {\bm \eta}_i(t)   ,
\end{align}
where  $\boldsymbol{\eta}_i(t)$
is a Gaussian white-noise random vector, with the moments
\begin{align}
\langle\boldsymbol{\eta} _i \rangle = &  \ 0  , \\
\langle {\eta}_{\alpha,i}(t)  {\eta}_{\beta,j}(t')\rangle = & \ 2 D_{r} \delta_{\alpha \beta} \delta_{ij} \delta(t-t').
\end{align}
Equation (\ref{eq:orient}) is a stochastic equation with multiplicative noise. An equivalent representation for polar coordiantes with additive noise is outlined in App. \ref{app:rotation}.
The translational and rotational diffusion ($D_{r}$) coefficients of a sphere are related via $D_t^0=D_{r}\sigma^2/3$. The importance of noise on the active-particle motion is characterized by  the P\'eclet number
\begin{align}
Pe=\frac{v_0}{\sigma D_r}  .
\end{align}

We study systems in the range of $9 < Pe < 300$. Up to $N=1.5\times 10^{5}$ particles are simulated in a cubic box of length $L$. The density is measured in terms of the
global packing fraction $\phi=\pi \sigma^3 N/(6V)$, where $V= L^3$ is the volume. A
natural time scale is the rotational relaxation time
$\tau_r=1/(2D_r)$. For the potential (\ref{eq:pot_energy}), we use the parameters $\sigma/l=60$ and $r_c=1.083 \sigma$.

\section{Dynamics of Active Brownian Particles} \label{sec:dynamics}

\subsection{Unconfined Particles at Infinite Dilution --- Diffusion} \label{sec:unconf_dilute}

The equations of motion for a single, force-free particle in an infinite or periodic system follows from  Eq.~(\ref{eq:BDtrans}) with $\bm F_i=0$.  Multiplication of the remaining  equation by ${\bm r}_i(t)$ leads to
\begin{align} \label{eq:averages_dilute}
\frac{1}{2}  \frac{d}{dt} \lla {\bm r}_i(t)^2 \rra = v_0 \lla {\bm e}_i(t) \cdot {\bm r}_i(t) \rra + \frac{1}{\gamma}  \lla{\bm \Gamma}_i (t) \cdot {\bm r}_i(t) \rra ,
\end{align}
where, $\lla \ldots \rra$ denotes the ensemble average. In the asymptotic limit $t \to \infty$, the average on the left-hand side is the mean  square displacement
\begin{align} \label{eq:msd}
\lla {\bm r}_i(t)^2 \rra = \lla {\bm r}_i(0)^2 \rra + 6 D_t^{id} t
\end{align}
for the homogeneous and isotropic system.  To evaluate the terms on the right-hand side, we use the formal solution
\begin{align} \label{eq:form_solution}
{\bm r}_i(t) = {\bm r}_i (-\infty) + \int_{-\infty}^{t}\left(v_0{\bm e}_i(t') + \frac{1}{\gamma} {\bm \Gamma}_i(t') \right) dt'
\end{align}
of Eq.~(\ref{eq:BDtrans}). Since ${\bm \eta}_i$ and ${\bm \Gamma}_i$ are independent stochastic processes, the terms on the right-hand side of Eq.~(\ref{eq:averages_dilute}) reduce to
\begin{align}  \label{eq:or_pos_corr}
\lla {\bm e}_i(t) \cdot {\bm r}_i(t) \rra & = v_0 \int_{-\infty}^t \lla {\bm e}_i(t)\cdot {\bm e}_i (t') \rra dt'  = \frac{v_0}{2 D_r} , \\ \label{eq:ran_pos_corr}
 \lla{\bm \Gamma}_i (t) \cdot {\bm r}_i(t) \rra  & = \frac{1}{\gamma} \int_{-\infty}^t \lla {\bm \Gamma}_i(t) \cdot {\bm \Gamma}_i(t') \rra dt' = 3 k_B T .
\end{align}
Here, we use the correlation function (cf. App.~\ref{app:correlation})
\begin{align}
\lla {\bm e}_i(t) \cdot {\bm e}_i(0) \rra = e^{-2D_rt} ,
\end{align}
and the definition
 \begin{align}
 \int_{-\infty}^{t} \delta(\hat t - t') dt' = \left\{
 \begin{array}{cc} \displaystyle
\displaystyle  1  , & \hat t \in (-\infty,t) \\[3pt] \displaystyle
 \frac{\displaystyle 1}{ \displaystyle 2} ,  &  \hat t = t \\[5pt]
 \displaystyle 0 , & \hat t  \in (t, \infty)
 \end{array}   \right. .
 \end{align}
Hence, Eq.~(\ref{eq:averages_dilute}) yields the diffusion coefficient
 \begin{align} \label{eq:diff_coef_active}
 D^{id}_t = D_t^0 + \frac{v_0^2}{6 D_r}
 \end{align}
 in the asymptotic limit $t\to \infty$. This is the well-know relation for the mean square displacement of an active Brownian particle in three dimensions.\cite{elge:15} However, we have derived the expression exploiting the virial theorem. \cite{beck:67} More general, in $d$ dimensions follows $D_t^{id} = D_t^0 + v_0^2/d(d-1) D_r$. \cite{elge:15} The expression has been  confirmed experimentally for 2D synthetic microswimmers in Refs.~\onlinecite{hows:07,volp:11}.

\subsection{Confined Particles --- Mean Particle Velocity}

For ABPs confined in a volume with impenetrable walls, the force on particle $i$ is
\begin{align} \label{eq:volume_force}
{\bm F}_i(t) = \psum_{j =1}^N  {\bm F}_{ij}(t) + {\bm F}_i^s(t) ,
\end{align}
where the ${\bm F}_{ij} = {\bm F}_{ij}({\bm r}_i-{\bm r}_j)$ are pairwise interparticle forces and ${\bm F}_i^s$ is the short-range  repulsive force with the wall. We assume that the wall force points along the local surface normal ${\bm n}$, i.e., ${\bm F}_i^s= - F_i^s {\bm n}_i$, where ${\bm n}_i = {\bm n}({\bm r}_i)$ points outward of the volume.   The prime at the sum of Eq.~(\ref{eq:volume_force}) indicates that the index $j=i$ is excluded.

By multiplying Eq.~(\ref{eq:BDtrans}) with the orientation vector $\bm e_i(t)$ and averaging over the random forces, we obtain
\begin{align} \label{eq:virial_orientation}
\sum_{i=1}^N \lla \dot {\bm r}_i \cdot \bm e_i \rra = N v_0 + \frac{1}{\gamma} \sum_{i=1}^N \lla \bm F_i \cdot  \bm e_i \rra .
\end{align}
The expression
\begin{align}
v = \frac{1}{N} \sum_{i=1}^N \lla \dot {\bm r}_i \cdot \bm e_i \rra
\end{align}
can be considered as the average particle velocity  along its propulsion direction in  the interacting system. \cite{solo:15} Hence, with Eqs.~(\ref{eq:volume_force}) and (\ref{eq:virial_orientation}), the propulsion velocity can be expressed as
\begin{align} \label{eq:average_velocity}
v =  v_0 + \frac{1}{\gamma N} \sum_{i=1}^N  \psum_{j =1}^N  \lla {\bm F}_{ij} \cdot \bm e_i \rra  + \frac{1}{\gamma N} \sum_{i=1}^N  \lla {\bm F}_i^s  \cdot  \bm e_i \rra .
\end{align}
Aside from the wall term, a similar expression has been derived in Ref.~\onlinecite{solo:15}. The pairwise interaction term can be written as
\begin{align} \label{eq:pair_orientation}
\sum_{i=1}^N  \psum_{j =1}^N  \lla {\bm F}_{ij} \cdot \bm e_i \rra = \frac{1}{2} \sum_{i=1}^N  \psum_{j =1}^N  \lla {\bm F}_{ij} \cdot \left( \bm e_i - \bm e_j \right) \rra .
\end{align}
As a consequence, we obtain a contribution to the velocity only when $\bm e_i \nparallel \bm e_j$.

The velocity perpendicular to a wall of a particle interacting with a wall is zero. Hence, the force $\lla F_i^s \rra$ of such a particle, averaged over the random force, is
\begin{align}
\lla F_i^s \rra = \gamma v_0 \lla \bm e_i \cdot \bm n_i \rra
\end{align}
for a dilute system. Thus,  the wall force yields a contribution to the mean propulsion velocity as along as $\bm e_i \not\perp \bm n_i$. In case of a dilute system, the mean velocity reduces then to
\begin{align} \label{eq:mean_prop_single}
v = v_0 - \frac{1}{\gamma N} \sum_{i=1}^N \lla F^s_i \bm e_i \cdot \bm n_i \rra = \frac{v_0}{N} \sum_{i=1}^N \left(1 - \lla [\bm e_i \cdot \bm n_i]^2 \rra \right) .
\end{align}
Hence, the presence of walls reduces the mean propulsion velocity. This is in accord with the considerations in Ref.~\onlinecite{solo:15.1} for anisotropic particles, but an extension of the calculations of Ref.~\onlinecite{solo:15} for spherical ABPs. Naturally, the velocity is zero for particles adsorbed at a wall and an orientation vector parallel to $\bm n$.

Generally, the average propulsion velocity of a particle, which is part time in the bulk, with probability $P_{bulk}$, and part time at the wall, with probability $1-P_{bulk}$,  is
\begin{align}
v_i = v_0 P_{bulk} + v_0 \left(1 - \lla (\bm e_i \cdot \bm n_i)^2 \rra \right) (1-P_{bulk}) .
\end{align}
Thus, the distribution function of the angle between $\bm e_i$ and $\bm n_i$ is required, as well as the probability distribution to find a particle in the bulk to determine the average propulsion velocity. Naturally, the average in Eq.~(\ref{eq:mean_prop_single}) can be calculated by the distribution function of the particle position and the propulsion orientation ${\bm e}_i$, which follows from the respective Fokker-Planck equation with appropriate boundary conditions.\cite{elge:13.1,solo:15}     \\

More general, multiplication of Eq.~(\ref{eq:orient}) by the velocity $\dot {\bm r}_i(0)$ yields
\begin{align}
\lla \bm e_i(t) \cdot \dot {\bm r}_i(0) \rra = \lla \bm e_i(0) \cdot \dot {\bm r}_i(0) \rra e^{-2D_rt} ,
\end{align}
within the Ito interpretation of the stochastic integral. \cite{risk:89,gard:83,raib:04} Then,  averaging over all particles gives
\begin{align} \label{eq:mean_velocity_correlation}
\frac{1}{N} \sum_{i=1}^N  \lla \bm e_i(t) \cdot \dot {\bm r}_i(0) \rra = v e^{-2D_rt} .
\end{align}
This relation describes the average decay of the correlation between the initial velocities $\dot {\bm r}_i(0)$ and the time-dependent orientations $\bm e_i(t)$. Not surprisingly, the correlation function decays exponentially, because the dynamics of the orientation $\bm e_i(t)$ is independent of the particle position and any particle interaction.

\section{Virial Formulation of Pressure} \label{sec:virial}

The expression of the pressure depends on the boundary conditions. Here, we will discuss a system of ABPs confined in  a cubic simulation box with either impenetrable walls or periodic boundary conditions.

\subsection{Confined System}

Multiplication of the equations of motion (\ref{eq:BDtrans}), with the forces (\ref{eq:volume_force}), by ${\bm r}_i(t)$  yields
\begin{align} \label{eq:basic_volume}
\gamma v_0 \lla {\bm e}_i(t) \cdot {\bm r}_i(t) \rra +   \lla{\bm \Gamma}_i (t) \cdot {\bm r}_i(t) \rra + \lla {\bm F}_i \cdot {\bm r}_i\rra =0 ,
\end{align}
since the mean square displacement $\lla {\bm r}_i(t)^2 \rra$ is limited for the confined system, and hence, its derivative vanishes in the stationary state. The force contribution can be written as
\begin{align} \label{eq:virial_volume}
\sum_{i=1}^N\lla {\bm F}_i \cdot {\bm r}_i\rra =  \frac{1}{2} \sum_{i}^N \psum_{j=1}^N \lla {\bm F}_{ij} \cdot ({\bm r}_{i} - {\bm r}_j) \rra + \sum_{i=1}^N \lla {\bm F}_i^s \cdot {\bm r}_i \rra.
\end{align}
The first term on the right-hand side is the virial of the internal forces and the second term that of the wall forces. The latter needs to be related to the pressure of the system, which is the major step in linking the mechanical (wall) pressure to bulk properties. The traditional derivation for equilibrium systems is briefly outlined in App. \ref{app:equi_pressure}. This derivation assumes a homogeneous  pressure throughout the volume. In contrast, active systems can exhibit pressure inhomogeneities depending on the shape of the confining wall, \cite{fily:14,smal:15} and thus, the derivation does not apply in general. So far, there is no general derivation or proof of the equivalence of the external virial and the pressure; there may be no such universal equivalence, since there is no thermodynamic potential for an active system.

However, for special geometries, we can establish a relation between the wall forces and pressure. This particularly applies to cuboidal and spherical volumes. As shown in App.~\ref{app:surf_pressure}, for such geometries the average wall pressure is given by
\begin{align} \label{eq:press_surface}
3 p^e V = - \sum_{i=1}^N \lla {\bm F}_i^s \cdot {\bm r}_i \rra ,
\end{align}
where we introduce the external pressure $p^e$ to indicate its origin by the external forces.
As an important result, summation of Eq.~(\ref{eq:basic_volume}) over all particles  yields the relation between pressure and bulk properties
\begin{align} \label{eq:press_volume}
3 p^iV = 3 N k_BT & \  + \gamma v_0\sum_{i=1}^N \lla {\bm e}_i \cdot {\bm r}_i \rra  +  \frac{1}{2} \sum_{i=1}^N \psum_{j=1}^N \lla {\bm F}_{ij} \cdot ({\bm r}_{i} - {\bm r}_j) \rra  .
\end{align}
Naturally, $p^i \equiv p^e$. However, we introduce the internal pressure $p^i$ to stress the different nature in the pressure calculation by bulk properties.
The first term on the right-hand side of Eq.~(\ref{eq:press_volume}) is the ideal-gas contribution to the pressure originating from the random stochastic forces ${\bm \Gamma}_i$ (Eq.~(\ref{eq:ran_pos_corr})). Without such forces,  our system reduces to that considered in Ref.~\onlinecite{yang:14.2}.
The other two terms are the virial contributions by the active and interparticle forces. Following the nomenclature of Ref.~\onlinecite{taka:14}, we denote second term on the right-hand side as ''swim virial''. It is a single-particle self contribution to the pressure, in contrast to the virial contribution by the interparticle forces.\cite{taka:14} Hence, the swim virial resembles similarities with a contribution by an external field rather than interparticle forces.

The contributions to the swim virial can be expressed as
\begin{align} \label{eq:correlation_integral}
\lla \bm e _i (t) \cdot \bm r_i (t) \rra & = \int_{- \infty}^t \lla \bm e_i(t) \cdot \dot {\bm r}_i (t') \rra dt' = \int_{0}^{\infty} \lla \bm e_i(t') \cdot \dot {\bm r}_i (0) \rra dt'
\end{align}
in the stationary state and with $\lla \bm r_i(-\infty) \cdot \bm e_i(t) \rra =0$. Together with Eq.~(\ref{eq:mean_velocity_correlation}),  Eq.~(\ref{eq:press_volume}) can be expressed as
\begin{align} \label{eq:pressure_mean_velocity}
3 p^iV = 3 N k_BT + \frac{\gamma N v_0}{2 D_r} v + \frac{1}{2} \sum_{i=1}^N \psum_{j=1}^N \lla {\bm F}_{ij} \cdot ({\bm r}_{i} - {\bm r}_j) \rra
\end{align}
in terms of the mean propulsion velocity $v$ (Eq.~\ref{eq:average_velocity}). A similar expression is obtained in Ref.~\onlinecite{solo:15}, but in a very different way. However, in contrast to Ref.~\onlinecite{solo:15}, our mean velocity (Eq.~\ref{eq:average_velocity}) comprises interparticle as well as wall contributions. Hence, the wall interactions are not only manifested in the pressure, but appear additionally in correlation functions, i.e., $\lla \bm e_i(t) \cdot \bm r_i(t) \rra$. This a a major difference to passive systems, were $v_0$ and hence $v$ are zero.

In the case of zero interparticle forces, the pressure Eq.~(\ref{eq:pressure_mean_velocity}) reduces to
\begin{align}
3 p^iV = 3 N k_B T \left(1 + \frac{v_0^2}{6 D_r D_t^0} \right) + \sum_{i=1}^N \lla \bm F_i^s \cdot \bm e_i \rra .
\end{align}
 The first term on the right-hand side corresponds to the ideal (bulk) pressure
\begin{align} \label{eq:ideal_pressure}
p^{id} = \frac{N k_BT}{V} \left(1 + \frac{v_0^2}{6 D_r D_t^0} \right)
\end{align}
of an active system.\cite{taka:14,mall:14,solo:15,smal:15} The second term accounts for wall interactions, and is responsible for various phenomena such as wall accumulation. \cite{elge:13.1,fily:14,yang:14.2,smal:15}

 \subsection{Periodic Boundary Conditions}

In systems with periodic boundary conditions, there is no wall force $\bm F^s$, but  in addition to interactions between the $N$ ''real'' particles,  interactions with periodic images located at ${\bm r}_i + \brn$ appear.
For a cubic simulation box, the vector $\brn$ is given by
\begin{align}
\brn = {\bm n} V^{1/3} ,
\end{align}
with $n_{\alpha} \in \mathbb{Z}$. The generalization to a rectangular noncubic box is straightforward. For a pair-wise potential $U_{ij}$, the potential energy is then given by
\begin{align} \label{eq:potential}
U = \frac{1}{2} \sum_{i=1}^N\psum_{j=1}^N  \sum_{{\bm n}} U_{ij}({\bm r}_i-{\bm r}_j - \brn) .
\end{align}
In general, this involves infinitely many interactions. \cite{wink:92.1,wink:02.1} Here, we will assume short-range interactions only. This limits the values $n_{\alpha}$ to $n_{\alpha} \pm 1$ for a particle in the primary box, which  corresponds to an application of the minimum-image convention. \cite{alle:87} The forces in Eq.~(\ref{eq:BDtrans}) are then given by
\begin{align}
{\bm F}_i = \psum_{j=1}^N \sum_{{\bm n}} {\bm F}_{ij}({\bm r}_i - {\bm r}_j - \brn) .
\end{align}
Multiplication of Eq.~(\ref{eq:BDtrans}) by ${\bm r}_i$ and summation over all particles leads to the relation
\begin{align}
v_0\sum_{i=1}^N \lla {\bm e}_i \cdot {\bm r}_i \rra & + \frac{1}{2 \gamma} \sum_{i=1}^N \psum_{j=1}^N \sum_{{\bm n}} \lla {\bm F}_{ij}^{\bm n} \cdot ({\bm r}_i - {\bm r_j}) \rra\\ \nonumber &  +\frac{1}{\gamma} \sum_{i=1}^N \lla {\bm \Gamma}_i \cdot {\bm r}_i \rra - \frac{1}{2} \sum_{i=1}^N \frac{d}{dt} \lla {\bm r}_i^2 \rra = 0 ,
\end{align}
with the abbreviation ${\bm F}_{ij}({\bm r}_i - {\bm r}_j - \brn) = {\bm F}_{ij}^{\bm n}$. By following the trajectories of the particles through the infinite system, i.e., by not switching to a periodic image in the primary box, when a particle leaves that box, $\lla{\bm r}_i(t)^2\rra$ is equal to the mean square displacement of Eq.~(\ref{eq:msd}), but with a translational diffusion coefficient $D_t$ of the interacting ABP system. Moreover, Eq.~(\ref{eq:ran_pos_corr}) applies and, hence,
\begin{align} \label{eq:virial_periodic_zero}
3 \gamma N D_t = 3 N k_BT  + \gamma v_0 \sum_{i=1}^N \lla {\bm e}_i \cdot {\bm r}_i \rra   + \frac{1}{2} \sum_{i=1}^N \psum_{j=1}^N \sum_{{\bm n}} \lla {\bm F}_{ij}^{\bm n} \cdot ({\bm r}_i - {\bm r_j}) \rra .
\end{align}
In the force-free case, i.e., ${\bm F}_{ij}^{\bm n}=0$, this relation is identical with Eq.~(\ref{eq:diff_coef_active}).
More general,  Eq.~(\ref{eq:virial_periodic_zero}) provides the self-diffusion coefficient of active particles in a periodic system interacting by pair-wise forces. Exploiting Eqs.~(\ref{eq:mean_velocity_correlation}) and
(\ref{eq:correlation_integral}), the active virial can be expressed by the mean propulsion velocity $v$, and the diffusion coefficient becomes
\begin{align} \label{eq:diff_active_periodic}
D_t=D_t^0 + \frac{v_0v}{6 D_r} + \frac{1}{6 N\gamma}  \sum_{i=1}^N \psum_{j=1}^N \sum_{{\bm n}} \lla {\bm F}_{ij}^{\bm n} \cdot ({\bm r}_i - {\bm r_j}) \rra ,
\end{align}
with the mean propulsion velocity in a  periodic system
\begin{align} \label{eq:mean_velocity_periodic}
v= v_0 + \frac{1}{N \gamma}  \sum_{i=1}^N \psum_{j=1}^N \sum_{{\bm n}} \lla {\bm F}_{ij}^{\bm n} \cdot \bm e_i \rra .
\end{align}
After insertion of the velocity $v$, Eq.~(\ref{eq:diff_active_periodic}) provides three contributions to the diffusion coefficient: (i) the diffusion coefficient, Eq.~(\ref{eq:diff_coef_active}), of non-interacting active Brownian particles, (ii) a contribution due to correlations between the forces ${\bm F}_{ij}^{\bm n}$ and the orientations $\bm e_i$, and (iii) a contribution from the virial of the inter-particle forces. This clearly reflects the explicit dependence of the diffusion coefficient on the orientational dynamics of the particles.

In contrast to confined systems, the multiplication of the equations of motion by the particle positions does not directly provide an expression for the pressure of a periodic system. Here, additional steps are necessary.
To derive an expression for the pressure, we subtract the term
\begin{align} \label{eq:virial_ex}
\frac{1}{2} \sum_{i=1}^N \psum_{j=1}^N \sum_{{\bm n}} \lla {\bm F}_{ij}^{\bm n} \cdot {\bm R}_{\bm n} \rra
\end{align}
from both sides of Eq.~(\ref{eq:virial_periodic_zero}). Then, we are able to define an external ($p^e$) and internal ($p^i$) pressure of the periodic system according to
\begin{align} \label{eq:press_e}
3p^eV  = & \ 3 N k_BT  \frac{D}{D_t} -\frac{1}{2} \sum_{i=1}^N \psum_{j=1}^N \sum_{{\bm n}} \lla {\bm F}_{ij}^{\bm n} \cdot {\bm R}_{\bm n} \rra  , \\ \label{eq:press_i}
3 p^i V  = & \ 3 N k_BT  + \gamma v_0 \sum_{i=1}^N \lla {\bm e}_i \cdot {\bm r}_i \rra   + \frac{1}{2} \sum_{i=1}^N \psum_{j=1}^N \sum_{{\bm n}} \lla {\bm F}_{ij}^{\bm n} \cdot ({\bm r}_i - {\bm r_j}-{\bm R}_{\bm n}) \rra ,
\end{align}
as for passive systems in Refs.~\onlinecite{wink:92.1,wink:93,wink:09}. Naturally, both expressions yield the same pressure value, i.e., $p^e \equiv p^i$.

For notational convenience, we write the expressions for the pressure as
\begin{align} \label{eq:press_e_app}
p^e & = p^d +p^{ef}  , \\ \label{eq:press_i_app}
p^i & = p^0 + p^s + p^{if} ,
\end{align}
with the abbreviations for the individual terms of Eqs.~(\ref{eq:press_e}) and (\ref{eq:press_i}):
\begin{align} \label{eq:press_0}
p^0 = &  \frac{N k_B T}{V} , \\ \label{eq:press_act}
p^s = & \frac{\gamma v_0}{3 V}  \sum_{i=1}^N \lla {\bm e}_i \cdot {\bm r}_i \rra =\frac{\gamma N}{6 V D_r} v_0 v  , \\ \label{eq:press_diff}
p^d = & \frac{N k_B T}{V}  \frac{D_t}{D_t^0}  ,\\ \label{eq:press_if}
p^{if} = & \frac{1}{6 V} \sum_{i=1}^N \psum_{j=1}^N \sum_{{\bm n}} \lla {\bm F}_{ij}^{\bm n} \cdot ({\bm r}_i - {\bm r_j}-{\bm R}_{\bm n}) \rra , \\ \label{eq:press_ef}
p^{ef} = & -\frac{1}{6V} \sum_{i=1}^N \psum_{j=1}^N \sum_{{\bm n}} \lla {\bm F}_{ij}^{\bm n} \cdot {\bm R}_{\bm n} \rra .
\end{align}

The expressions Eqs.~(\ref{eq:press_e}) and (\ref{eq:press_i}) are extensions of those of a conservative passive system, where $D_t=0$ and $v_0=0$, with a similar definition of external and internal pressure. \cite{wink:92.1,wink:92.2}  For both cases, the definitions appear {\em  ad hoc}. Note that for a passive system typically Newtons' equations of motion are considered. However, for the passive system, the expression for the internal pressure can be derived from the free energy,\cite{gree:47} as illustrated in the appendix of Ref.~\onlinecite{wink:92.2}, which underlines the usefulness and correctness of the applied procedure for passive systems.  The similarity between the internal pressure expressions Eqs.~(\ref{eq:press_volume}) and (\ref{eq:press_i}) for the confined and periodic systems, respectively, supports the adequateness of the applied procedure in the pressure calculation.

The individual terms can be understood as follows.

{\em External pressure $p^e$ ---}
The pressure contribution $p^{ef}$  accounts formally for the forces across faces of the periodic system, similar to the expression (\ref{eq:press_surface}) of a confined system. There is only a contribution for non-zero vectors ${\bm R}_{\bm n}$, i.e., for interactions between real and image particles.\cite{wink:92.1}
The diffusive dynamics yields the pressure contribution $p^d \sim D_t/D_t^0$.  This term is not present in a conservative passive system and is a consequence of the stochastic forces $\bm \Gamma_i$ . \cite{wink:92.1,wink:93,wink:09} For passive, non-interacting  particles, i.e., $v_0 =0$ and $U$ of Eq.~(\ref{eq:potential}) is zero, $D_t= D_t^0$ and the pressure reduces to the ideal gas expression $p^e =p^0$. For non-interacting but active systems, $D_t$ is given by Eq.~(\ref{eq:diff_coef_active}) and $p^e = p^{id}$  (Eq.~(\ref{eq:ideal_pressure})).\\

{\em Internal pressure $p^i$ ---}
The internal pressure comprises the ideal gas contribution $p^0$, the swim pressure $p^s$ due to activity, and the contribution $p^{if}$ by  all the pair-wise interactions in the system within the minimum-image convention.  As for the virial of the confined system (\ref{eq:pressure_mean_velocity}), the swim pressure $p^s$ Eq.~(\ref{eq:press_act}) can be expressed by the mean propulsion velocity $v$ (Eq.~(\ref{eq:mean_velocity_periodic})).
The forces ${\bm F}_{ij}^{\bm n}$ can be strongly correlated with the propagation direction ${\bm e}_i$, specifically when the system phase separates as in Ref.~\onlinecite{wyso:14}. For non-interacting particles or at very low concentrations, the contribution of the intermolecular forces vanishes and the pressure is given by Eq.~(\ref{eq:ideal_pressure}).\\

There are various similarities between the expressions for the pressure of a confined and a periodic system. Formally, Eqs.~(\ref{eq:pressure_mean_velocity}) and (\ref{eq:press_i}) are similar. However, specifically the meaning of the mean propulsion velocity $v$ is very different. The presence of the wall term in Eq.~(\ref{eq:average_velocity}) can lead to wall-induced phenomena, which are not present in periodic systems. Hence, results of simulations of periodic and confined active systems  are not necessarily identical.

\begin{figure}[t]
\includegraphics*[width=\columnwidth]{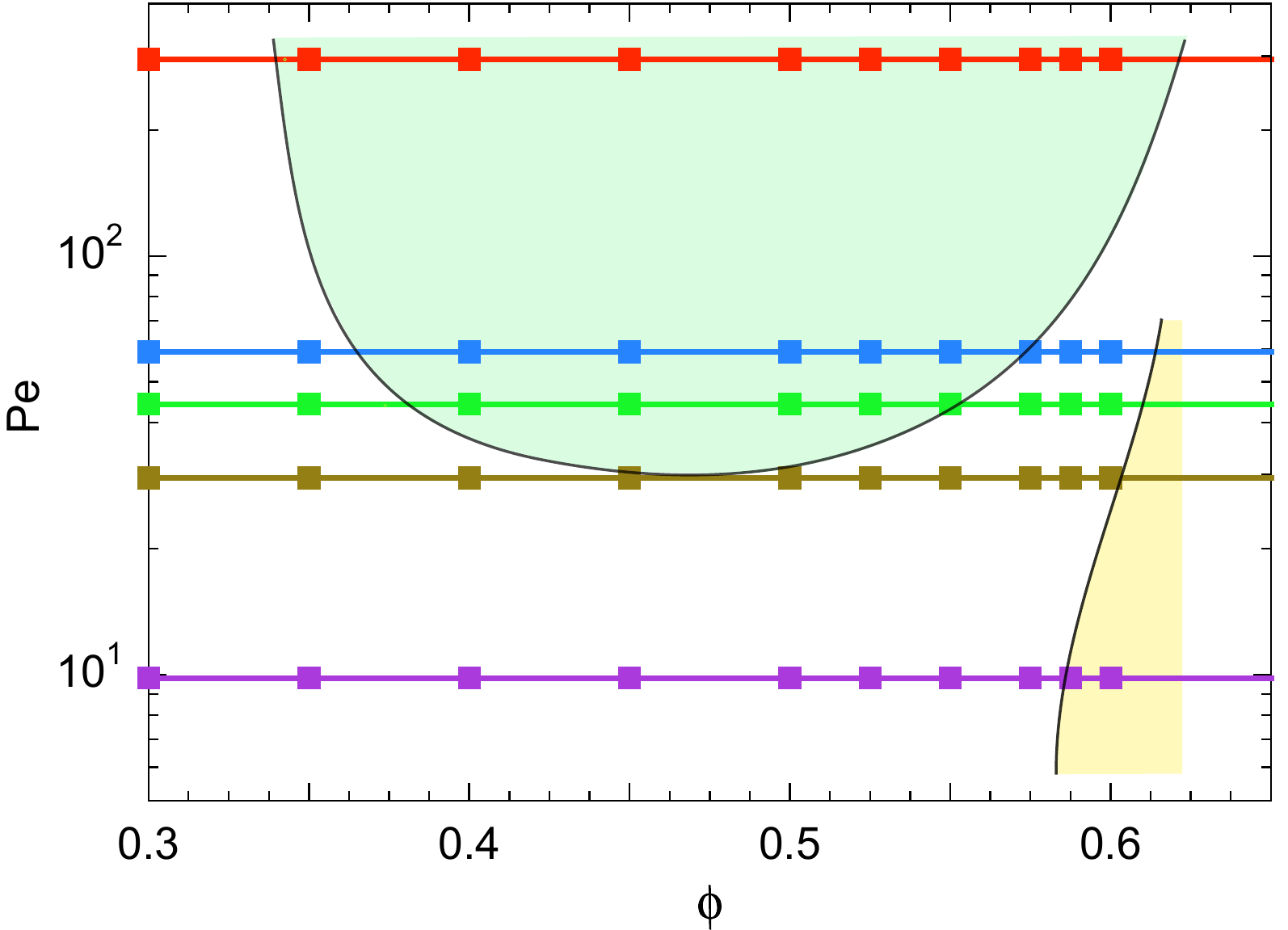}
\caption{Schematic phase diagram of active Brownian particles in terms of the P\'eclet number $Pe$ and the packing fraction  $\phi$. Symbols correspond to the P\'eclet numbers for which the pressure is calculated. The shaded areas indicate the liquid-gas (light green) and the crystal-gas (light yellow) coexistence regimes. Further details can be found in Ref.~\protect\onlinecite{wyso:14}. } \label{fig:phase_diagram}
\end{figure}

\section{Simulation Results} \label{sec:simulation}

Simulations of active Brownian spheres with periodic boundary conditions  (cf. Sec.~\ref{sec:model}) reveal an intriguing phase behavior and a highly collective dynamics, as discussed in Ref.~\onlinecite{wyso:14} for three dimensional systems. At low concentrations, the ABPs exhibit a uniform gas-like phase. However, above a critical, P\'eclet-number-dependent concentration, the isotropic system phase separates into a dilute, gas-like phase, and a dense, liquid-like phase \cite{wyso:14,sten:14}. Figure~\ref{fig:phase_diagram} displays the phase diagram for the considered concentrations and P\'{e}clet numbers. Above $Pe \approx 30$, we find a wide range of concentrations, over which the system phase separates.

As is evident from Eqs.~(\ref{eq:press_e}) and (\ref{eq:press_i}), diffusion contributes to the pressure. Hence, we address the dynamics of the ABPs in terms of the diffusion coefficient in the following. Subsequently, we discuss the dependence of the pressure and its individual contributions on P\'eclet number and concentration.

\begin{figure}[t]
\includegraphics*[width=\columnwidth]{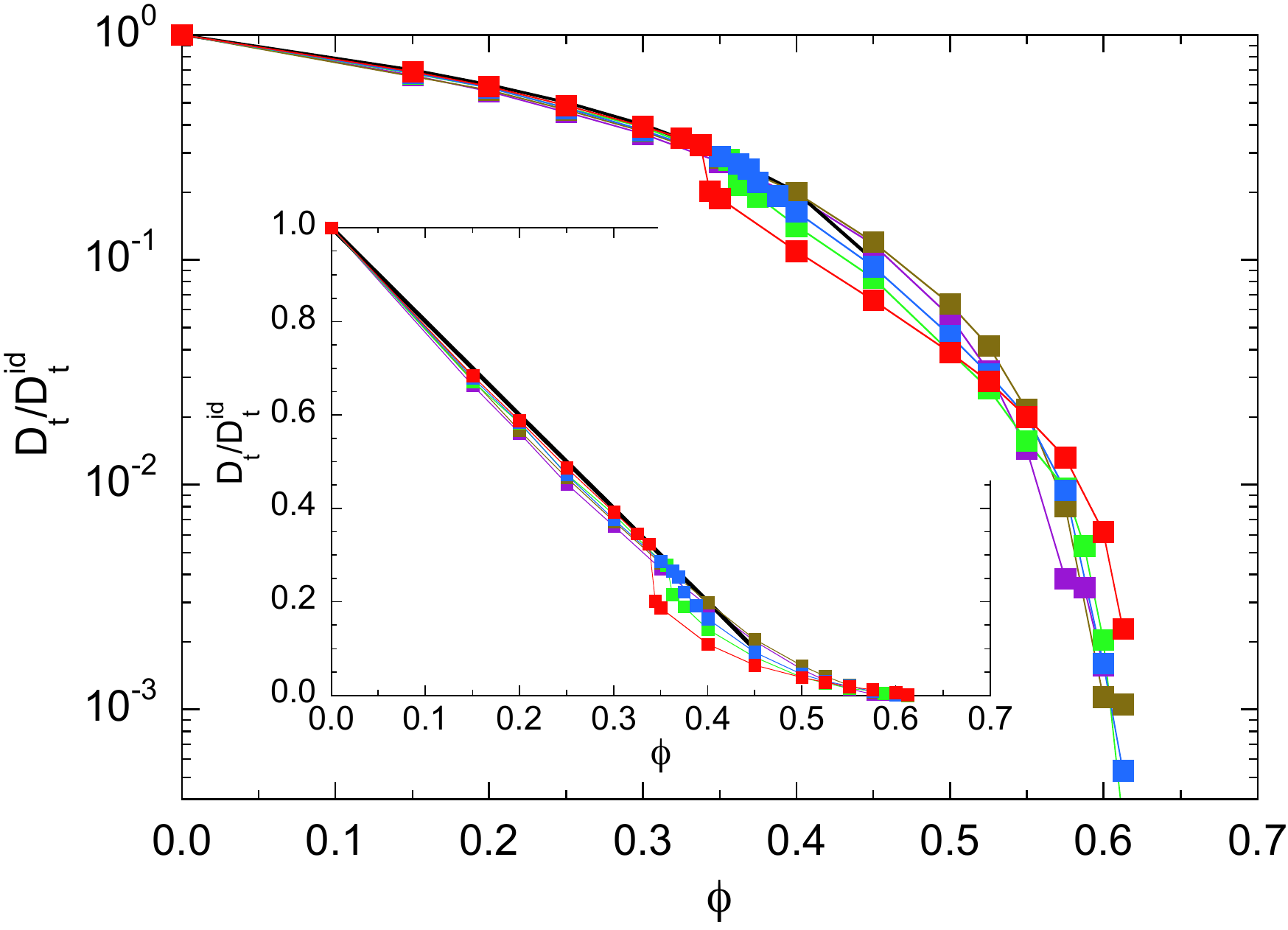}
\caption{Translational diffusion coefficients of ABPs for the P\'eclet numbers $Pe=9.8$ (purple), $29.5$ (olive), $44.3$ (green), $59.0$ (blue), and $295.0$ (red). The values are normalized by the respective diffusion coefficients (\ref{eq:diff_coef_active}) of an individual ABP. The inset shows the same data on linear scales. The solid line (black) indicates the dependence $D_t/D_t^{id} = 1-2\phi$. } \label{fig:diffusion_density}
\end{figure}

\subsection{Diffusion coefficient of ABPs}

We determine the diffusion coefficient of the ABPs by the Einstein relation, i.e., we calculated the particle mean square displacement and extracted from the linear long-time behavior the translational coefficient $D_t$. Figure~\ref{fig:diffusion_density} displays the obtained values as function of the packing fraction for various P\'eclet numbers. For small packing fractions ($\phi \lesssim 0.3$), $D_t/D_t^{id}$ decreases linearly with increasing concentration. The decay is reasonably well described by the theoretically expected relation $1-2\phi$ for colloids in solution.\cite{dhon:96} However, a better agreement is obtained for the relation $1-2.2\phi$, i.e., a $10\%$ large factor. Remarkably, the decrease of $D_t$ for $\phi \lesssim 0.3$ is essentially independent of the activity. Significant deviations between the individual curves are visible for higher concentrations and larger P\'eclet numbers, where a phase separation appears. At the critical concentration, the diffusion coefficients for $Pe>30$ exhibit a rapid change to smaller values. A further increase in  concentration implies a further slowdown of the dynamics. Interestingly, at high concentrations, the larger P\'eclet number ABPs diffuse faster than those with a small $Pe$. We attribute this observation to collective effects of the fluids exhibiting swirl- and jet-like structures. \cite{wyso:14} Another reason might by an activity induced fluidization, since the glass transition is shifted to larger packing fractions.\cite{ni:13}

\begin{figure}[t]
\includegraphics*[width=\columnwidth]{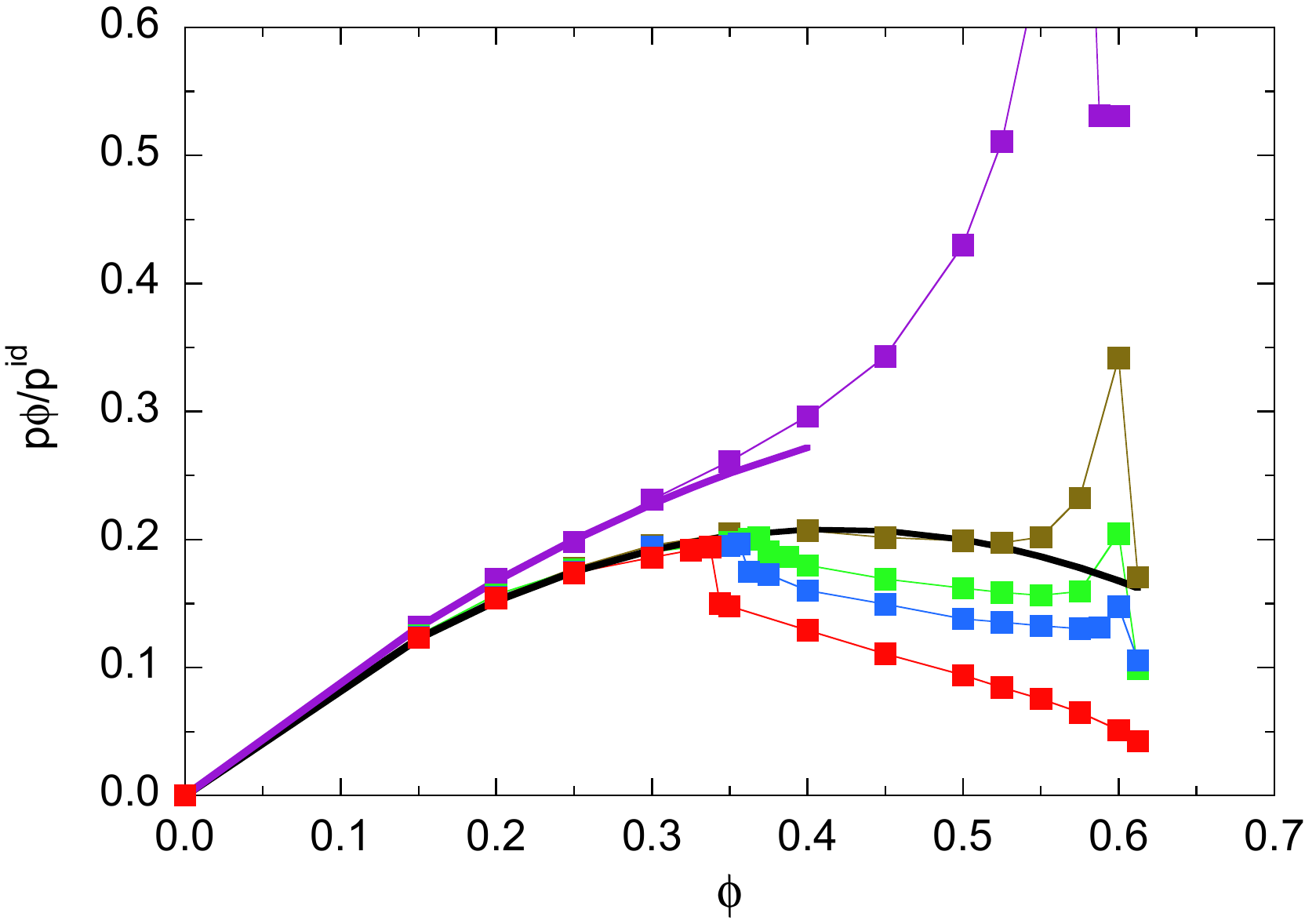}
\caption{Pressure in systems of ABPs for the P\'eclet numbers $Pe=9.8$ (purple), $29.5$ (olive), $44.3$ (green), $59.0$ (blue), and $295.0$ (red). The values are normalized by the pressure $p^{id}$ (\ref{eq:ideal_pressure}) of individual ABPs at the same P\'eclet number.  The black line indicates the dependence $p\phi/p^{id} = \phi(1-1.2\phi)$, and the purple line the relation $p\phi/p^{id} = \phi(1-0.8\phi)$.} \label{fig:pressure_tot}
\end{figure}

\begin{figure}[t]
\includegraphics*[width=\columnwidth]{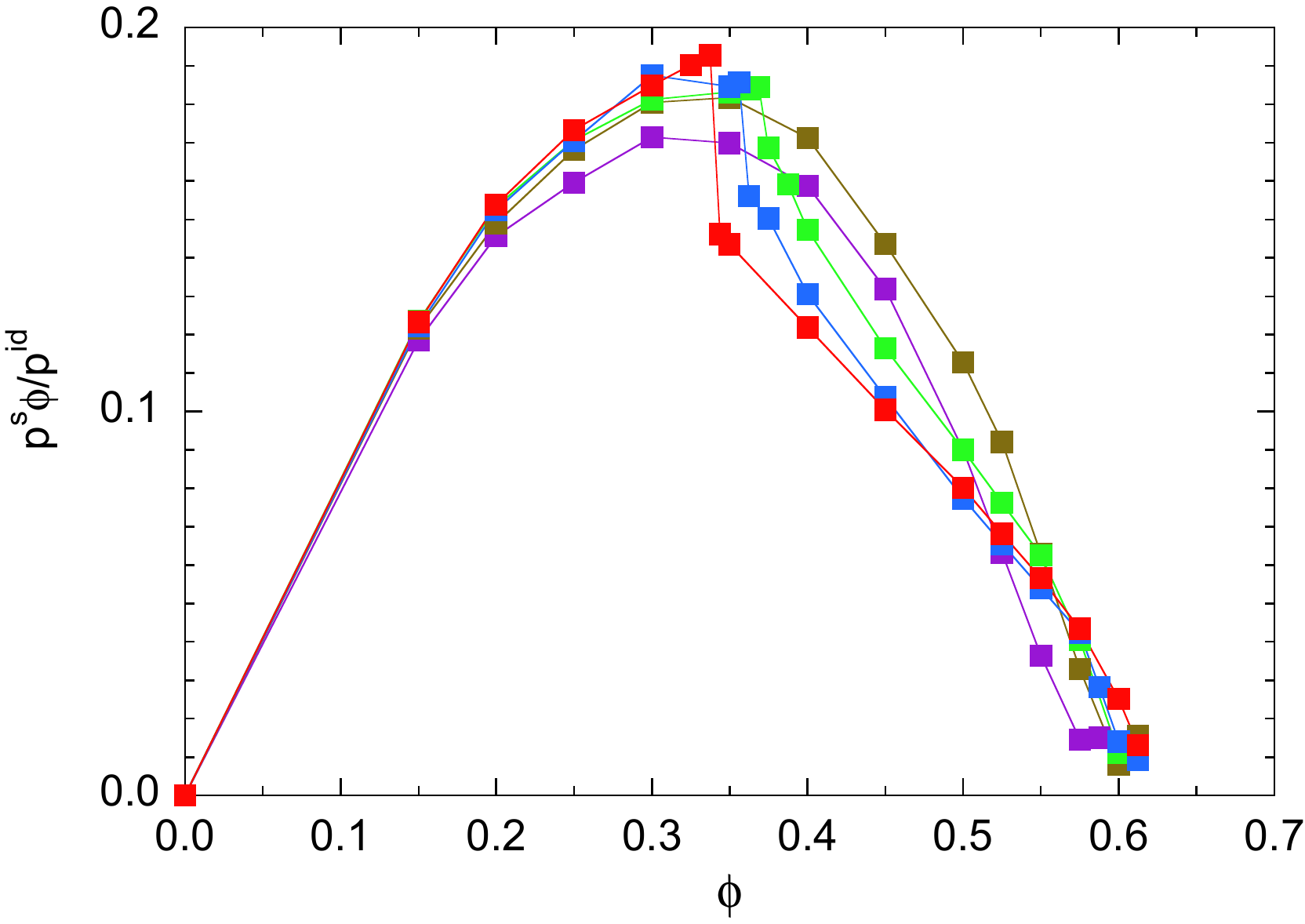}
\caption{Active pressure Eq.~(\ref{eq:press_act}) in systems of ABPs for the P\'eclet numbers $Pe=9.8$ (purple), $29.5$ (olive), $44.3$ (green), $59.0$ (blue), and $295.0$ (red). The values are normalized by the pressure $p^{id}$ (\ref{eq:ideal_pressure}) of individual ABPs at the same P\'eclet number. } \label{fig:pressure_active}
\end{figure}

\begin{figure}[t]
\includegraphics*[width=\columnwidth]{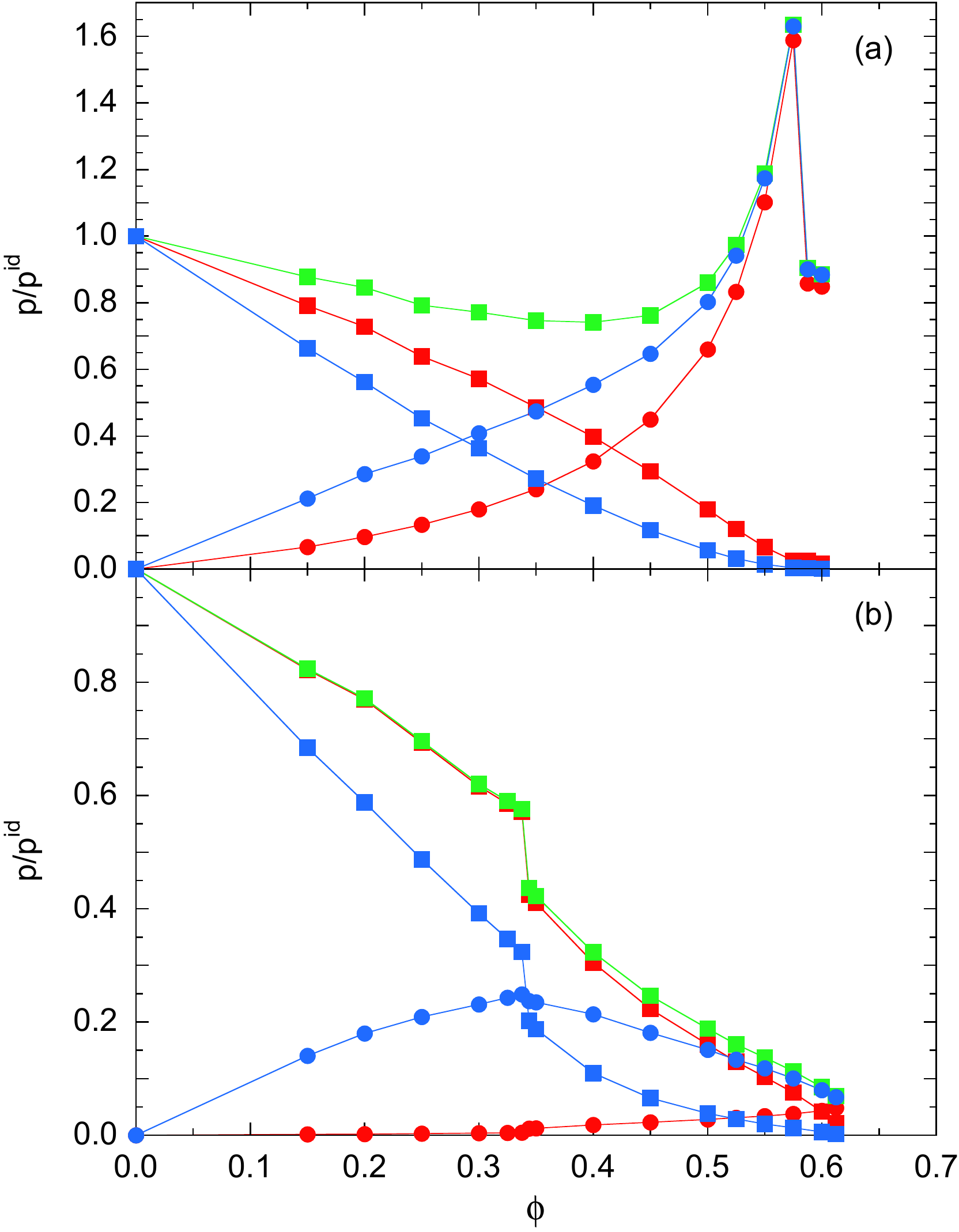}
\caption{Pressure and its individual contributions of ABPs for the P\'eclet number (a) $Pe=9.8$  and (b) $Pe=295.0$. The values are normalized by the pressure $p^{id}$ (\ref{eq:ideal_pressure}) of individual ABPs at the same P\'eclet number. The green symbols indicate the total pressure Eqs.~(\ref{eq:press_e_app}) and (\ref{eq:press_i_app}). The red symbols correspond to internal pressure components $p^s$ (${\color{red}\blacksquare}$)  and $p^{if}$ (${\color{red}\newmoon}$) of Eqs.~(\ref{eq:press_act}) and (\ref{eq:press_if}), respectively. The blue symbols indicate the pressure components $p^d$ (${\color{blue}\blacksquare}$)   and $p^{ef}$ (${\color{blue}\newmoon}$) of Eqs.~(\ref{eq:press_diff}) and (\ref{eq:press_ef}), respectively. } \label{fig:pressure_components}
\end{figure}

\begin{figure}[t]
\includegraphics*[width=\columnwidth]{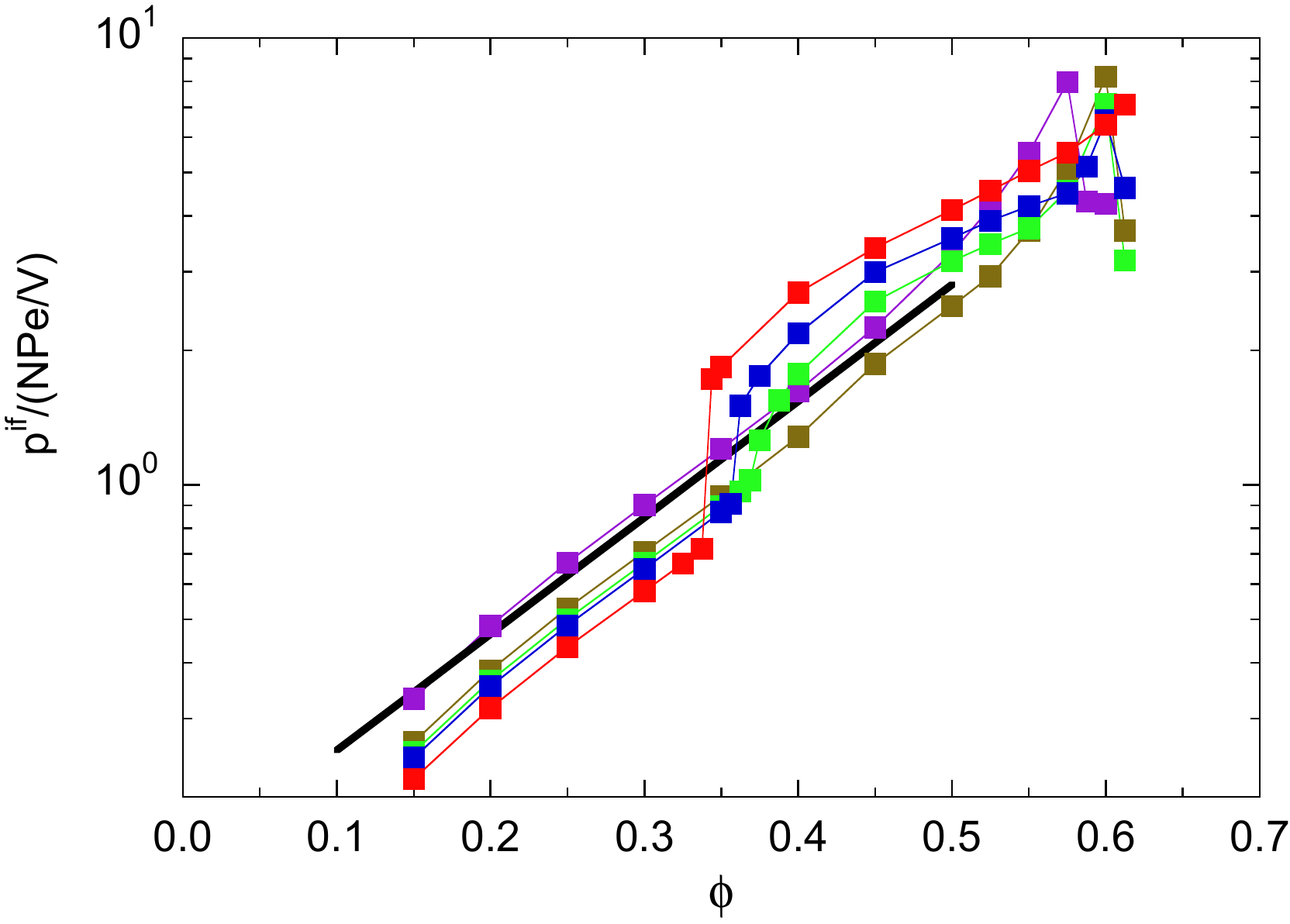}
\caption{Contribution of the interparticle force (Eq.~(\ref{eq:press_if})) to the internal pressure  of ABPs for the P\'eclet numbers $Pe=9.8$ (purple), $29.5$ (olive), $44.3$ (green), $59.0$ (blue), and $295.0$ (red).  The solid line (black) indicates the dependence $p^{if} \sim e^{6\phi}$. } \label{fig:pressure_force_internal}
\end{figure}

\begin{figure}[t]
\includegraphics*[width=\columnwidth]{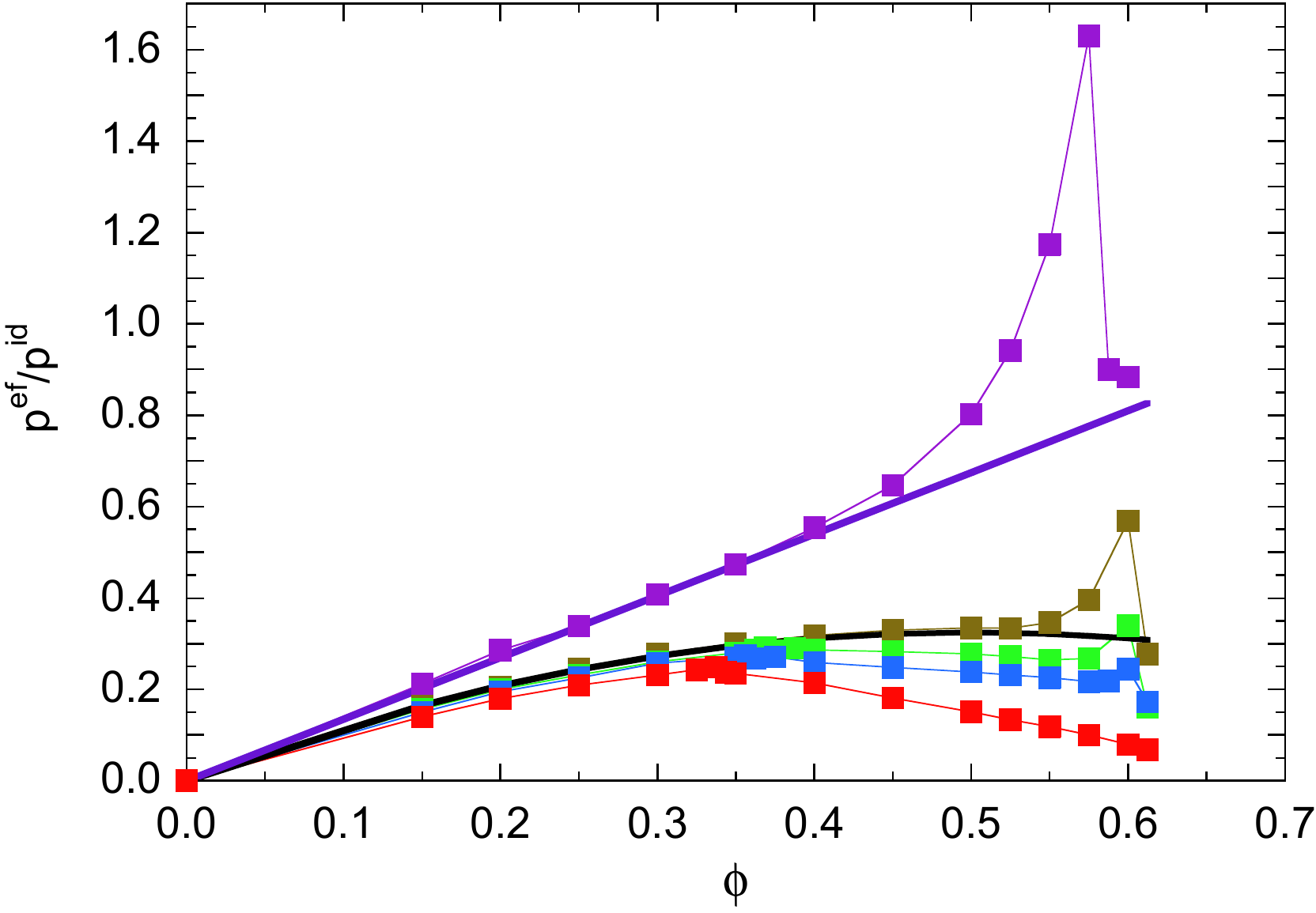}
\caption{Contribution of the interparticle force (Eq.~(\ref{eq:press_ef})) to the external pressure  of ABPs for the P\'eclet numbers $Pe=9.8$ (purple), $29.5$ (olive), $44.3$ (green), $59.0$ (blue), and $295.0$ (red).  The solid lines indicates the dependencies $p^{ef}/p^{id} \approx 1.35 \phi$ (purple) and $p^{ef}/p^{id} \approx 1.3 \phi(1-\phi)$ (black). } \label{fig:pressure_force_external}
\end{figure}

\subsection{Pressure of ABPs}

Figure~\ref{fig:pressure_tot} displays the density dependence of the  total pressure $p\equiv p^e \equiv p^1$ for the considered P\'eclet numbers. We normalize the pressure by the ideal gas value (\ref{eq:ideal_pressure}) and multiply by the packing fraction $\phi$ in order to retain the linear density dependence. As expected, both, Eq.~(\ref{eq:press_e_app}) and Eq.~(\ref{eq:press_i_app}) yield the same pressure values. We observe a pronounced dependence of the pressure on the P\'eclet number. Note that the pressure for the largest P\'eclet number is highest, because $p^{id} \sim Pe^2$ for $Pe \gg 1$. For small $Pe$, the pressure increases with increasing $\phi$. However, already for $Pe = 29.5$, we obtain a strongly nonmonotonic concentration dependence, where the pressure decreases again for $\phi \gtrsim 0.35$. A similar behavior has already been obtained in Refs.~\onlinecite{taka:14,yang:14.2,solo:15}. In Ref.~\onlinecite{taka:14}, the decreasing pressure at high concentrations is attributed to cluster formation of the ABPs.
For ABPs with P\'eclet numbers well in the phase separating region, we find a jump of the pressure at the critical packing fraction. This behavior has not been observed before, most likely because
smaller P\'eclet numbers have been considered only. Such a pressure jump seems not to be present for $Pe=29.5$, but the Pe\'clet number is close to the binodal (cf. Fig.~\ref{fig:phase_diagram}) and a clear phase separation is difficult to detect in simulations.

Interestingly, all the pressure curves exhibit a remarkably similar behavior for $Pe \geqslant 29.5$ and $\phi< 0.35$ and can very well be fitted by the density dependence
\begin{align} \label{eq:nonlin_density}
\frac{p\phi}{p^{id}} = \phi (1-\kappa \phi)
\end{align}
in this regime, with $\kappa \approx 1.2$, as indicated in Fig.~\ref{fig:pressure_tot}. For smaller $Pe=9.8$, we find  $\kappa \approx 0.8$. Hence, in any case the pressure increases linearly for $\phi \ll 1$, as expected, and exhibits quadratic corrections for larger concentrations, which depend on the swimming velocity or P\'eclet number.
In Ref.~\onlinecite{taka:14}, the nonequilibrium virial equation of state
\begin{align} \label{eq:peos}
p =  p^{id} \left[ 1-(1-3/2Pe) \phi\right]
\end{align}
is proposed, adopted to our definition of the P\'eclet number, and $p^{id}= N k_BT(1-Pe^2/2)/V$. This relation applies only within a certain range of $Pe$ values. It certainly accounts for the increase of the prefactor of $\phi$ with increasing $Pe$, but the asymptotic value for $Pe \gg 1$ does not agree with our simulation result. However,  our simulation results  for $Pe=9.8$ are reasonably well described by Eq.~(\ref{eq:peos}) at small concentrations.

The nonlinear concentration dependence of Eq.~(\ref{eq:nonlin_density}) with a positive $\kappa$ yields a negative second-virial coefficient $B_2=-\pi \sigma^3 \kappa /6$. In classical thermodynamics, a negative $B_2$ suggest the possibility of a gas-liquid phase transition. \cite{taka:14} For the ABPs, the negative $B_2$ definitely indicates large concentration fluctuations, which increase with increasing P\'eclet number. However,  in the light of our simulation results, it seems that a phase transition is also associated with a pressure jump at the critical concentration.
This is more clearly expressed by the swim pressure $p^s$ displayed in Fig.~\ref{fig:pressure_active}. Only for $Pe>29.5$, we find a pressure jump and a phase transition.

Figure~\ref{fig:pressure_components} illustrates the contributions of the individual terms of Eqs.~(\ref{eq:press_act}) -- (\ref{eq:press_ef}) to the total pressure for the P\'eclet numbers $Pe=9.8$ and $Pe=295$. At small $Pe = 9.8$ (see Fig.~\ref{fig:pressure_components}(a)), the terms $p^s$ and $p^d$ decrease monotonically with increasing concentration and approach zero at large $\phi$.  For $p^d$, this is easily understood by the slowdown of the diffusive dynamics due to an increasing frequency of particle encounters with increasing density.   The pressure $p^d$ is given by
\begin{align} \label{eq:p_act_p_diff}
p^d = p^0 + p^s + \frac{1}{6V}  \sum_{i=1}^N \psum_{j=1}^N \sum_{{\bm n}} \lla {\bm F}_{ij}^{\bm n} \cdot ({\bm r}_i - {\bm r_j}) \rra
\end{align}
in terms of the ideal gas and the swim pressure. As for a passive system, the last term on the right-hand side of this equation is smaller than zero, and hence, $p^s > p^d-p^0$. The equation also shows that the swim pressure is generally different from $p^d$ and hence is not simply related to diffusive behavior of the ABPs.
As far as the force contributions  $p^{ef}$ and $p^{if}$ are concerned, they monotonically increase with $\phi$ for $Pe=9.8$, until the ABPs undergo a phase transition and form a crystalline structure. The gradual increase of pairwise interactions with increasing concentration leads to a systematic increase of this pressure contribution. The ratio $p/p^{id}$ decreases initial ($\phi < 0.4$) due to the slowdown of the diffusive dynamics, but at large $\phi$, the contributions $p^{ef}$ and $p^{if}$ dominate and the respective pressure increases with concentration similar to that of a passive system. For $Pe=9.8$, $p$ is essentially equal to  $p^{ef}$ for $\phi > 0.5$.

The relevance of the individual contributions changes with increasing P\'eclet number. For $Pe=295$ (see Fig.~\ref{fig:pressure_components}(b)), $p^d$ and $p^s$ still decrease monotonically with increasing $\phi$. However, the total pressure is essentially identical with $p^s$, i.e., $p$ is dominated and determined by the swim pressure. The contribution $p^{if}$ due to interparticle interactions is very small and is only relevant at high concentrations. The pressure $p^{ef}$ increase for small concentrations, passes through a maximum and decreases again for large $\phi$. This nonmonotonic behavior is very different from that at low-P\'eclet numbers. The appearing jumps in $p^s$ and $p^d$ reflect the appearing phase transition at a critical density $\phi_c$.\cite{wyso:14} Note that the critical density depends on the P\'eclet number.

Aside from absolute values, the density dependence of the various pressure components is rather similar for densities below the critical density $\phi_c \approx 0.35$. In any case, the pressure is dominated by the active contribution as expressed and reflected in the diffusion coefficient. Above  $\phi_c$, however, the contribution by the inter-particle forces dominates for $Pe=9.8$, whereas $p$ is still determined by $p^s$ for $Pe=295$. Here, the contribution of $p^s$ exceeds that of $p^{if}$ by far, although the latter increases also with the P\'eclet number as shown in Fig.~\ref{fig:pressure_force_internal}, but $p^s \sim Pe^2$ at small $\phi$ and outweighes the growth of $p^{if}$. Two effects contribute to the decrease of the overall pressure. The diffusion coefficient decreases with increasing concentration due to an increasing number of particle-particle encounters. However, $p^d$ decreases faster than $p^s$. Via the relation (\ref{eq:p_act_p_diff}), we see that this faster decrease is compensated by an increase in the contribution by the interparticle forces.

Looking at the swim pressure $p^s$ (cf. Eq.~(\ref{eq:mean_velocity_periodic})), it evidently decreases because the correlations between the forces and the orientations $\bm e_i$ decrease with increasing concentration. Here, an important aspect is the appearance of collective effects of the ABPs, specifically for $\phi > \phi_c$. \cite{wyso:14} Collective effects lead to a preferential parallel alignment of certain nearby ABPs. As indicated in Eq.~(\ref{eq:pair_orientation}), such an alignment reduces the force-propulsion-direction correlations, and hence, the pressure.

As indicated before, the force contribution to the internal pressure $p^{if}$ depends roughly linearly on $Pe$, as illustrated in Fig.~\ref{fig:pressure_force_internal}. Remarkably, $p^{if}$ exhibits an exponential dependence on density, namely $p^{if} \sim \exp(6 \phi)$, over a wide range of concentrations for P\'eclet numbers in the one-phase regime, and up to the critical value $\phi_c$ for P\'eclet numbers where a phase transition appears. Additionally, the forces exhibit a rapid increase at the critical density. We did not analyse in detail whether a jump appears or if $p^{if}$ is smooth and a gradual but rapid increase occurs, as might be expected from the data for $Pe=44.3$.

Figure~\ref{fig:pressure_force_external} displays the contributions of the internal forces (Eq.~(\ref{eq:press_ef})) to the external pressure (Eq.~(\ref{eq:press_e_app})). The general shape of the curves is comparable to that of the scaled pressure $p\phi/p^{id}$. Compared to the latter, $p^{ef}$ exhibits the stronger density dependence $p^{ef} \sim \phi^2$ for $\phi \ll 1$. However, compared to $p^{if}$, the density dependence is weak. Quantitatively, it can be described as $p^{ef}/p^{id} = \hat \kappa \phi(1-\phi)$ with $\hat \kappa \approx 1.35$ over a wide range of P\'eclet numbers $9 < Pe \lesssim 30$. With increasing $Pe$, $\hat \kappa$ decreases slightly and we find $\hat \kappa \approx 1.1$ for $Pe=295$.

\section{Summary and Conclusions} \label{sec:conclusions}

We have presented theoretical and simulation results for the pressure in systems of spherical active Brownian particles in three dimensions. In a first step, we have shown that the mechanical pressure as force per area can be expressed via the viral by bulk properties for particular geometries, especially a cuboid volume as typically used in computer simulations. Subsequently, we have derived expressions for the pressure via the virial theorem of systems confined in cuboidal volumes or with periodic boundary conditions. For the latter, two equivalent expressions are derived, denoted as internal and external pressure, respectively. As a novel aspect, we considered overdamped equations of motion for active systems in the presence of a Gaussian and Markovian white-noise source. The latter gives rise to a thermal diffusive dynamics for the translational motion and yields the ideal-gas contribution to the pressure.

The activity of the Brownian particles yields distinct contributions to the respective expressions of the pressure. For the external pressure, an effective diffusion coefficient of an active particle appears. The internal pressure contains a term, which is related to the product of the bare propulsion velocity of an ABP and its effective mean propulsion velocity. The latter is determined by the interactions with surrounding ABPs and reduces to the bare velocity at infinite dilution. Similar expressions have been derived in Refs.~\cite{taka:14,yang:14.2,solo:15}

Our simulations of ABPs in systems with three-dimensional periodic boundary conditions, show that the total pressure increases first with concentration and decreases then again at higher concentrations for P\'eclet numbers exceeding a ceratin value, in agreement with previous studies. \cite{taka:14,yang:14.2} Thereby, the pressure exhibits a van der Waals-like pressure-concentration curve. This shape clearly reflects collective effects, however, it does not necessarily indicate a phase transition.  In case of a phase transition, we find a rapid and apparently discontinues reduction of the pressure at the phase-separation density.

Quantitatively, the density dependence of the pressure is well described by the relation $p \sim \phi(1-\kappa\phi)$, where the range over which the formula applies and $\kappa$  depends weakly on the P\'eclet number. For $Pe\approx 10$, we find $\kappa \approx 0.8$. For P\'eclet numbers in the vicinity and above $Pe_c$, all the considered curves are well described with $\kappa \approx 1.2$. Generally speaking, our $\kappa$ values extracted from the simulations are in the vicinity of the value $\kappa=1$ proposed in Ref.~\onlinecite{taka:14}.

Considering the various contributions to the pressure individually, we find that the internal pressure is dominated by the contribution from activity for $Pe> Pe_c$, in particular for $\phi < \phi_c$.  For P\'eclet numbers larger than the critical value, there is significant contribution from the interparticle forces of Eq.~(\ref{eq:mean_velocity_periodic}). This contribution decreases with increasing P\'eclet number, since the swim pressure increase quadratically with the P\'eclet number and the pressure due to interparticle force linearly. However, for moderate $Pe$, the exponential increase of $p^{if}$ with concentration outweighes the P\'eclet-number dependence.

In summary, the pressure of an active system exhibits distinctive differences to a passive system during an activity-induced transition from a (dilute) fluid to a system with a coexisting gas and a dense liquid phase. While the pressure of a passive system remains unchanged during the phase transition, our simulations of an active system yield a rapid decrease of the pressure for densities in the vicinity of the phase-transition density. Thereby, the pressure change becomes more pronounced with increasing P\'eclet number. Hence, evaluation of the pressure, specifically the swim pressure, provides valuable insight into the phase behavior of active systems. \\

We gratefully acknowledge support from the DFG within the priority program SPP 1726 ''Microswimmers''.


\begin{appendix}
\section{Rotational Equation of Motion} \label{app:rotation}

Equation (\ref{eq:orient}) is a stochastic equation with multiplicative noise. The Fokker-Planck equation, which yields the correct equilibrium distribution function, follows with the white noise in the Stratonovich calculus. \cite{risk:89,lyut:14,raib:04}

For the numerical integration of the equations of motion (\ref{eq:orient}), an Ito representation is more desirable. Introducing spherical coordinates according to
\begin{align}
{\bm e} =
\left( \begin{array}{c}
\cos \varphi \ \sin \vartheta \\
\sin \varphi \ \sin \vartheta \\
\cos  \vartheta \\
\end{array} \right)
\end{align}
(here, we consider a particular particle and skip the index $i$), alternatively the stochastic differential equations can be considered
\begin{align} \label{eq:app_diff_angles}
\dot {\bm e}= \eta_{\vartheta} {\bm e}_{\vartheta}+ \eta_{\varphi} {\bm e}_{\varphi}  ,
\end{align}
and yields the same Fokker-Planck equation for the generalized coordinates ${\varphi}$ and $\vartheta$. \cite{raib:04}  Thereby, the equations of motion of the angles are
\begin{align} \label{eq:app_diff_theta}
\dot \vartheta & = D \cot \vartheta +  \eta_{\vartheta} \ , \\ \label{eq:app_diff_phi}
\dot \varphi & = \frac{1}{\sin \vartheta}  \eta_{\vartheta} \ ,
\end{align}
with additive noise.
The unit vectors ${\bm e}_{\xi}$, $\xi \in \{\varphi, \vartheta\}$,  follow by differentiation ${\bm e}_{\xi} \sim \partial {\bm e}/\partial \xi$ and normalization. The $\eta_{\xi}$ are again Gaussian and Markovian random processes with the moments
\begin{align} \label{eq:app_noise_differential} \nonumber
\lla \eta_{\xi} \rra = &  \ 0  , \\
\lla {\eta}_{\xi}(t)  {\eta}_{\xi'}(t') \rra = & \ 2 D_{r} \delta_{\xi \xi'}  \delta(t-t').
\end{align}

For simulation, Eq.~(\ref{eq:app_diff_angles}) has to be discretized. Integration, iteration, and usage of the Eqs.~(\ref{eq:app_diff_theta}) and (\ref{eq:app_diff_phi}) yields the difference equation
\begin{align} \label{eq:app_discrete}
\bm e(t+\Delta t) =  \bm e(t)+ {\bm e}_{\vartheta}(t) \Delta \eta_{\vartheta}  + {\bm e}_{\varphi}(t) \Delta\eta_{\varphi} - 2 D_r \bm e(t) \Delta t.
\end{align}
The $\Delta \eta_\xi$ are defined as
\begin{align}
\Delta \eta_\xi = \int_t^{t+\Delta t} \eta_\xi(t') dt'
\end{align}
and are Gaussian random numbers with the first two moments
\begin{align} \nonumber \label{eq:app_noise_discrete}
\lla \Delta \eta_\xi \rra & =  \ 0  \\
\lla \Delta \eta_{\xi} \Delta \eta_{\xi'} \rra & =  2 D_r \delta_{\xi \xi'} \Delta t .
\end{align}
This is easily show with the help of Eqs.~(\ref{eq:app_noise_differential}).\cite{gard:83} As common for the Ito calculus, we replaced the values $\Delta \eta_{\xi}^2$ by their averages $2 D_r \Delta t$ in Eq.~(\ref{eq:app_discrete}). \cite{gard:83,raib:04}

For practical reasons, it is more convenient to apply the following integration scheme. An ''estimate'' $\bm e'(t+\Delta t)$ is obtained via
\begin{align} \label{eq:app_discrete_approx}
\bm e'(t+\Delta t) =  \bm e(t)+ {\bm e}_{\vartheta}(t) \Delta \eta_{\vartheta}  + {\bm e}_{\varphi}(t) \Delta\eta_{\varphi} .
\end{align}
This new orientation vector is no longer normalized. Normalization yields
$\bm e (t+\Delta t) = \bm e'(t+\Delta t) / |\bm e'(t+\Delta t)|$,  where
 $|\bm e'(t+\Delta t)|^2= \bm e'(t+\Delta t)\cdot\bm e'(t+\Delta t) = 1+ \Delta \eta_{\vartheta}^2  + \Delta\eta_{\varphi}^2$. Since $\Delta\eta_{\varphi}^2 \sim \Delta t$, we obtain for $D_r \Delta t \ll 1$
\begin{align} \label{eq:app_discrete_random} \nonumber
\bm e(t+\Delta t)  =  \bm e(t) & + {\bm e}_{\vartheta}(t) \Delta \eta_{\vartheta}  + {\bm e}_{\varphi}(t) \Delta\eta_{\varphi} \\ &
 - \frac{1}{2} \left(\Delta \eta_{\vartheta}^2  + \Delta\eta_{\varphi}^2 \right)\bm e(t)
\end{align}
up to order $\Delta t^{3/2}$.  Upon replacement of $\Delta \eta_{\xi}^2$ by its average, we obtain the desired equation (\ref{eq:app_discrete}). \\

\section{Orientational Vector Correlation Function} \label{app:correlation}

The autocorrelation function of the orientational vector is conveniently obtained from Eq.~(\ref{eq:orient}) for an infinitesimal time interval $dt$, i.e., the equation
\begin{align}
d\bm e = {\bm e}(t) \times d {\bm \eta}(t)  -2 D_r \bm e dt
\end{align}
within the Ito calculus.
Multiplication with $\bm e(0)$ and averaging yields
\begin{align}
\lla d \bm e(t) \cdot {\bm e}(0) \rra   = &  \lla \left[{\bm e}(t) \times d {\bm \eta}(t) \right]\cdot \bm e(0) \rra
 - 2 D_r  \lla \bm e(t) \cdot \bm e(0) \rra d t .
\end{align}
Since $\bm e(t)$ is an nonanticipating function, \cite{gard:83} the average over the stochastic-force term vanishes, and we get
\begin{align}  \nonumber
\frac{d}{dt} \lla \bm e(t) \cdot {\bm e}(0) \rra   = - 2 D_r  \lla \bm e(t) \cdot \bm e(0) \rra ,
\end{align}
which gives
\begin{align}  \nonumber
 \lla \bm e(t) \cdot {\bm e}(0) \rra   = e^{- 2 D_r t} .
\end{align}

\section{Virial and Pressure in Equilibrium Systems} \label{app:equi_pressure}

To relate the external virial with the pressure for an equilibrium system, we consider a surface element $\Delta S$ and the forces exerted on that surface element $\sum_{i \in \Delta S} \bm F_i^s$. The pressure $p^e$ at the surface element is
\begin{align}
p^e \Delta S = \sum_{i \in \Delta S} \lla F_i^s \rra = - \sum_{i \in \Delta S} \lla \bm F_i^s \bm n_i \rra.
\end{align}
On the other hand, the external virial reads as
\begin{align} \label{eq:surf_virial_app}
\sum_{i \in \Delta S} \lla {\bm F}_i^s \cdot {\bm r}_i \rra  = \lla {\bm r} \cdot \sum_{i \in \Delta S}  {\bm F}_i^s \rra .
\end{align}
Since the surface element is small, the vector $\bm r_i$ is essentially the same for all particles interacting with $\Delta S$ and can be replaced by the vector $\bm r$ of that element. The same applies to the normal $\bm n_i$, i.e., $\bm n_i = \bm n$, where $\bm n$ is the local normal at $\Delta S$.
With Eq.~(\ref{eq:surf_virial_app}) and the mechanical pressure, we then obtain
\begin{align}
 {\bm r} \cdot \sum_{i \in \Delta S} \lla  {\bm F}_i^s \rra = - {\bm r}  \cdot \bm n \sum_{i \in \Delta S}  \lla F_i^s \rra  =  - p^e {\bm r}  \cdot \Delta {\bm S} ,
\end{align}
with $\Delta \bm S = \bm n \Delta S$. Integration over the whole surface gives
\begin{align}
- \oint p^e \, {\bm r} \cdot d {\bm S} = \sum_{i=1}^N \lla {\bm F}_i^s \cdot {\bm r}_i \rra.
\end{align}
Using Gauss theorem, the surface integral turns into a volume integral, i.e.,
\begin{align}
\oint p^e \, {\bm r} \cdot d {\bm S} = \int \nabla \cdot (p \, {\bm r}) \ d^3 r = 3 p^e V .
\end{align}
For the last equality, we assume that the pressure is isotropic and homogeneous. Hence, \cite{beck:67}
\begin{align}
3p^eV = - \sum_{i=1}^N \lla {\bm F}_i^s \cdot {\bm r}_i \rra .
\end{align}

\section{Virial and Pressure of Active Particles} \label{app:surf_pressure}

\subsection{Cuboidal Confinement}

\begin{figure}[t]
\includegraphics*[width=0.8\columnwidth]{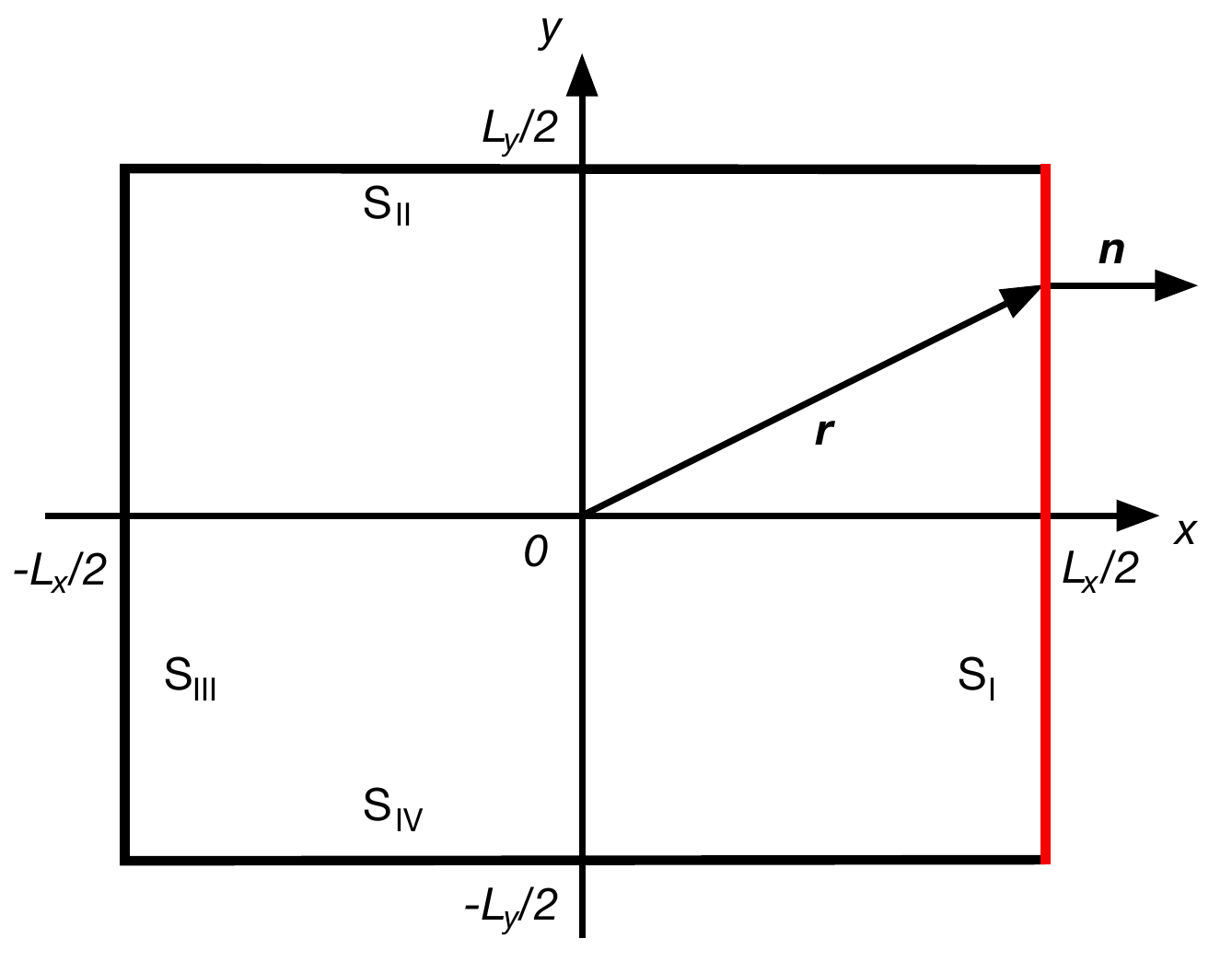}
\caption{Cross section of a cuboid for the calculation of the mechanical pressure.} \label{fig:cuboid}
\end{figure}

We consider particles confined in a cuboidal volume of dimensions $L_x\times L_y \times L_z$, with the reference frame located in its center (cf. Fig.~\ref{fig:cuboid}). The total mechanical pressure is given by the average over the six surfaces, i.e., $p^e = (p_I+p_{II} + \ldots + p_{VI})/6 $. For the surface $S_I$, the mechanical pressure is
\begin{align}
p_I = \frac{1}{S_I} \sum_i \lla F_i^s \rra .
\end{align}
Since $\bm n \cdot \bm r_i = L_x/2 $ for any point on the surface, we introduce unity, i.e., $ \bm n \cdot \bm r_i/(L_x/2) =1 $ and obtain
\begin{align}
p_I =   - \frac{2}{S_IL_x} \sum_i \lla {\bm F}_i^s \cdot \bm r_i  \rra=   - \frac{2}{V} \sum_i \lla {\bm F}_i^s \cdot \bm r_i \rra ,
\end{align}
with $\bm F_i^s =  - F_i^s \bm n $.
Hence, the pressure is given by
\begin{align} \label{app:surface_virial}
3 p^e V = - \sum_{i=1}^N \lla  {\bm F}_i^s \cdot \bm r_i \rra ,
\end{align}
as for an equilibrium system. The derivation of this relation is independent of the choice of the Cartesian reference frame. At least, as long as the axis unit-vectors are normal to the surfaces.

\subsection{Spherical Confinement}

For a sphere of radius $R$ with the reference frame in its center, the pressure is
\begin{align}
p^e = \frac{1}{S} \sum_{i=1}^N \lla F_i^s \rra= - \frac{1}{S} \sum_{i=1}^N \lla {\bm F}_i^s \cdot {\bm n} \rra,
\end{align}
where $S=4 \pi R^2$. For every point on the surface applies $\bm r_i = R \bm n$, hence,
\begin{align}
p^e = - \frac{1}{SR} \sum_{i=1}^N \lla {\bm F}_i^s \cdot {\bm r}_i \rra .
\end{align}
With $SR= 4\pi R^3 = 3V$, we again find the relation (\ref{app:surface_virial}).

\end{appendix}

\providecommand*{\mcitethebibliography}{\thebibliography}
\csname @ifundefined\endcsname{endmcitethebibliography}
{\let\endmcitethebibliography\endthebibliography}{}

\end{document}